\documentclass[slac_one]{revtex4}
\usepackage{graphicx}
\usepackage{fancyhdr}
\newcommand{\mett}{\mbox{$E\!\!\!\!/_{\rm T}$}}
\newcommand{\invfb}{$\rm fb^{-1}$}

\begin{document}

\title{Searches for New Physics at CDF} %% Paper title goes here


%\author{J. Physicist, R. N. P. Winner}
%\affiliation{SLAC, Stanford, CA 94025, USA}
%
\author{Max Goncharov for the CDF Collaboration}
\affiliation{Texas A\&M University, College Station, TX, USA}

\begin{abstract}
  In this note we present the results of several searches for physics
  beyond the standard model (SM). All final states use 2~\invfb\ of data 
  produced at Tevatron in $p\bar{p}$ collisions at $\sqrt{s}=1.96$~TeV and
  collected by the CDF Run II detector. None of the analysis is heavily optimized 
  for any specific model. All signatures contain 
  missing transverse energy (\mett) in the 
  final state and all, except for one, have at least one photon ($\gamma$) in
  the final state. Unfortunately, no interesting excesses of events over the 
  SM prediction are observed in any final states, except may be for the last one.
\end{abstract}

\maketitle

\thispagestyle{fancy}


\section{Introduction}

This note describes the results of CDF exotic searches in six final states:
di-jet~+~\mett, di-photon~+~\mett, photon~+~jets~+~\mett, 
photon~+~b-jet~+~jet~+~\mett, and photon~+~lepton~+~b-jet~+~\mett.
All final states, especially the first two, are sensitive to production cross-section 
enhancements due to gauge-mediated super-symmetry
(GMSB) models~\cite{gmsb}, while the last one, for example, is essentially $t\bar{t}\gamma$ 
final state that has never been looked into at CDF. All analysis either exploit 
new data driven techniques to estimate the SM backgrounds or make use of 
new hardware that became available after Run IIb upgrade.
 
The di-jet~+~\mett\ analysis uses data driven background estimates that 
were perfected in the monojet channel~\cite{monoPRL}. 
The di-photon~+~\mett\ and  photon~+~jets~+~\mett\ signatures make use of newly 
developed \mett\ model that allows to 
separate events with real \mett\ from events with fake \mett\ based 
on the extensive knowledge of energy fluctuations in the CDF detector.
They also use the EMTiming system~\cite{emtime}
to cut away or estimate non-collision photon candidates induced by cosmic rays and 
beam halo. The  photon~+~jets~+~\mett\ photon~+~b-jet~+~jet~+~\mett\ signatures 
use newly installed shower pre-radiator 
system~\cite{cp2} that allows to predict the fraction of pure photons in all photon
candidates.

\section{Exclusive Di-jet + \mett} % Section title should be in all capitals.

  Events with large \mett\ and one or more energetic jets can be produced in 
many models of new physics as well as in SM electroweak and 
QCD processes. The magnitude of the \mett\ and the number 
of jets depend on the specific model of new physics, while the SM backgrounds 
and instrumental effects can be studied independently.  
The base event sample is selected using kinematic requirements 
of event $H_{\rm T} > 125$~GeV~\cite{htdef} and $\mett > 80$~GeV.
We also perform a separate search in the high kinematic region defined by 
$H_{\rm T} > 225$~GeV and $\mett > 100$~GeV. 
In both regions, we compare the expected SM backgrounds with observed data.

\begin{figure*}[h]
\centering
\includegraphics[width=0.4\linewidth]{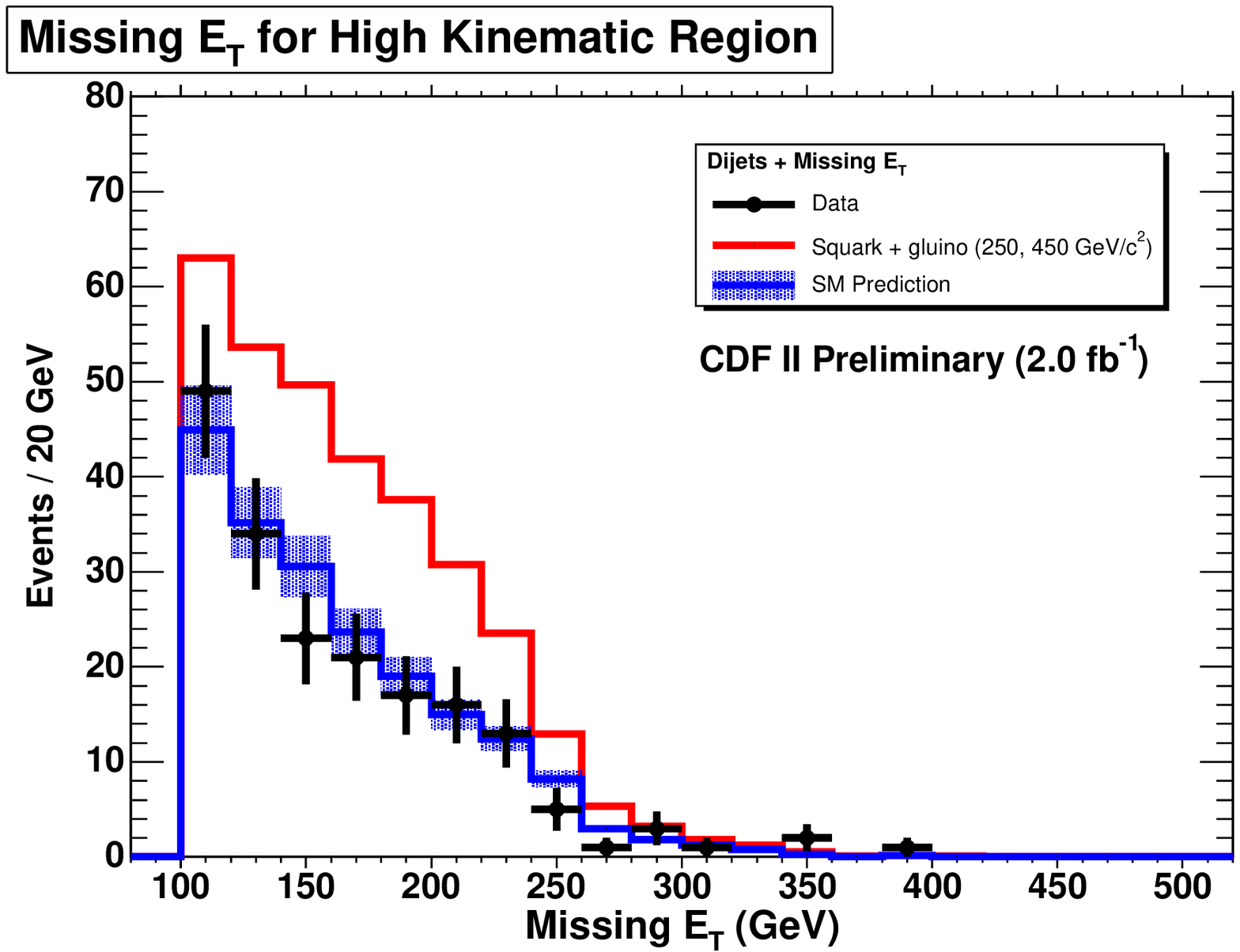}
\includegraphics[width=0.4\linewidth]{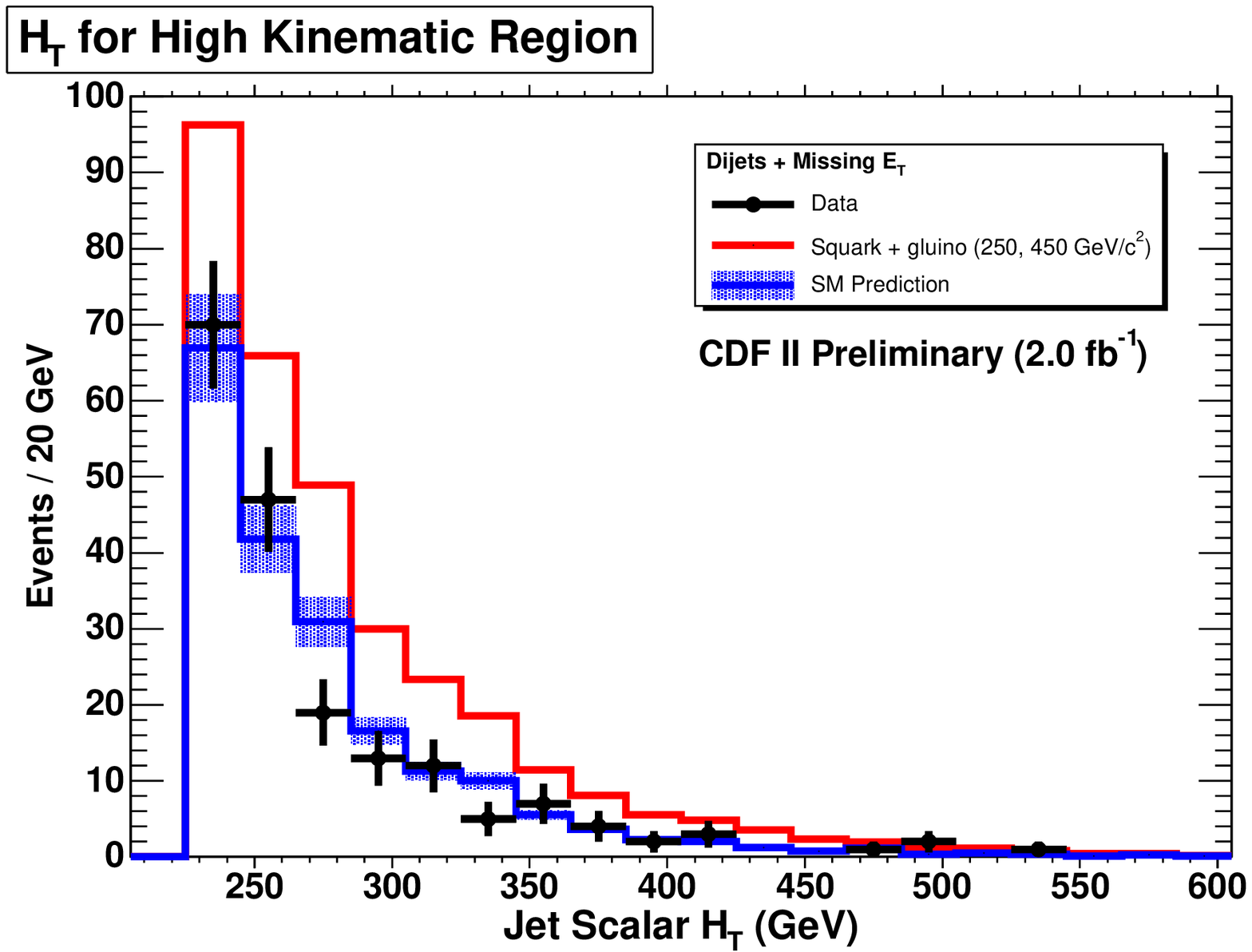}
\caption{Comparison of Standard Model Background Estimate with events 
observed in data in the high kinematic region as a function of event 
missing $E_{\rm T}$  and jet scalar $H_{\rm T}$ (linear scale). The additional potential 
signal contribution from MSSM spectrum S2 -- one of the points ruled out 
at leading order -- is also shown. Note that the high kinematic region is 
the more sensitive of the two for this mass spectrum.} 
\label{jjmet}
\end{figure*}

The SM estimate reveals $2312 \pm 140$ events from SM on top of  2506
events observed in data. For high kinematic region we find
$196 \pm 26$ events from SM and
186 events in data. The two leading backgrounds are 
$W+{\rm jet,}\ W\rightarrow l\nu$ ($102 \pm 10$) and
$Z+{\rm jet,}\ Z\rightarrow \nu\bar{\nu}$ ($71 \pm 12$) production.
We interpret the result in terms of cross-section limits on generic 
minimal super-symmetric (MSSM) models. Figure~\ref{jjmet} shows \mett\ and
$H_{\rm T}$ distributions, aw well as the prediction from MSSM model.

\section{Di-photon + \mett}

In this section we present a model-independent search for anomalous production of
$\gamma\gamma+\mett$ events. New \mett\ model is developed that uses
the best of our knowledge about energy fluctuations, or energy mis-measurements, 
in the CDF detector. The model provides the \mett\ probability distribution function 
for individual events and allows efficient separation of events with \mett\ due to
undetectable particles, like $\nu$ or $\tilde{G}$, from events where
\mett\ is caused by mis-measured jet energies.

\begin{figure*}[h]
\centering
\includegraphics[width=0.4\linewidth]{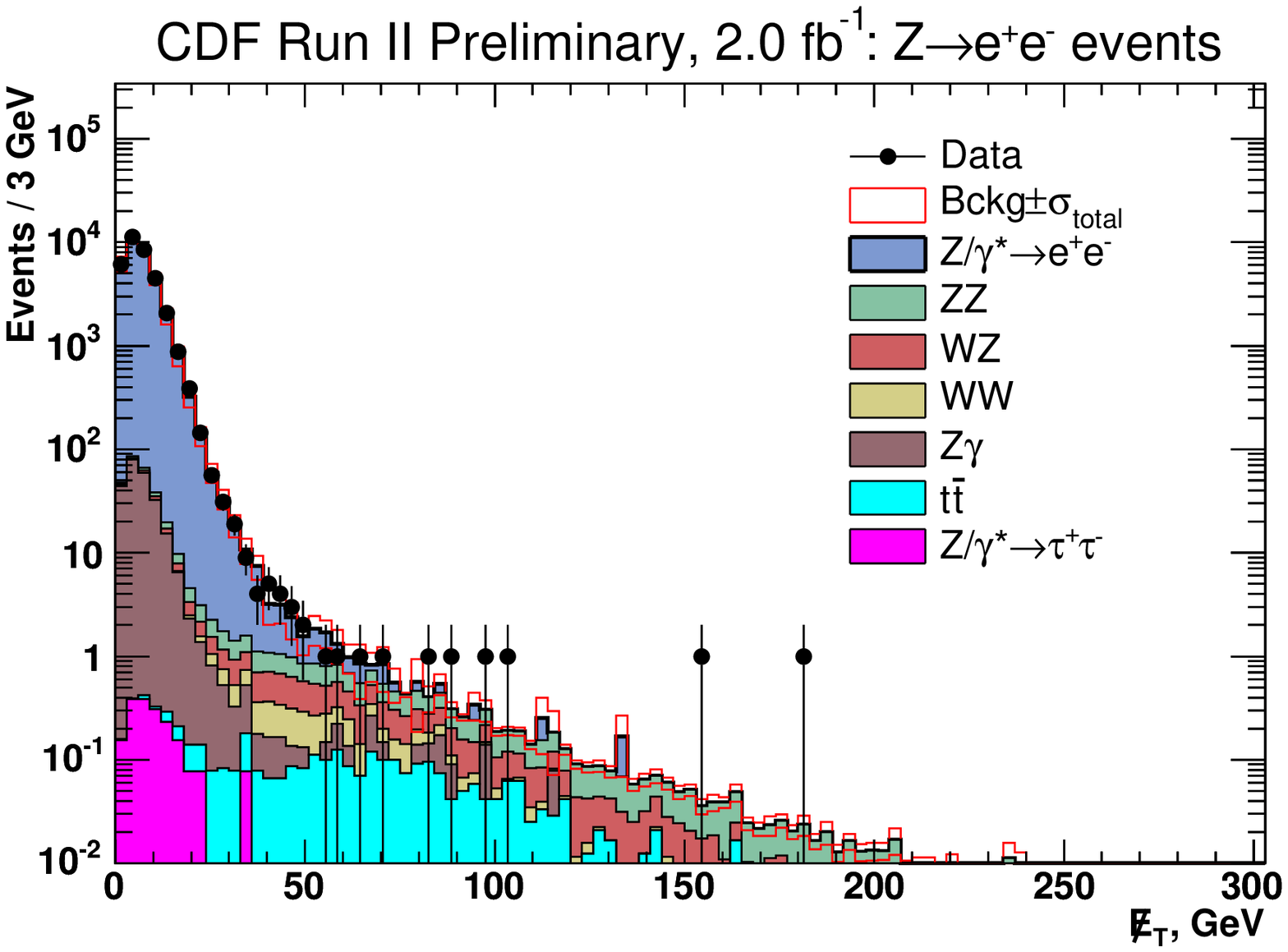}
\includegraphics[width=0.4\linewidth]{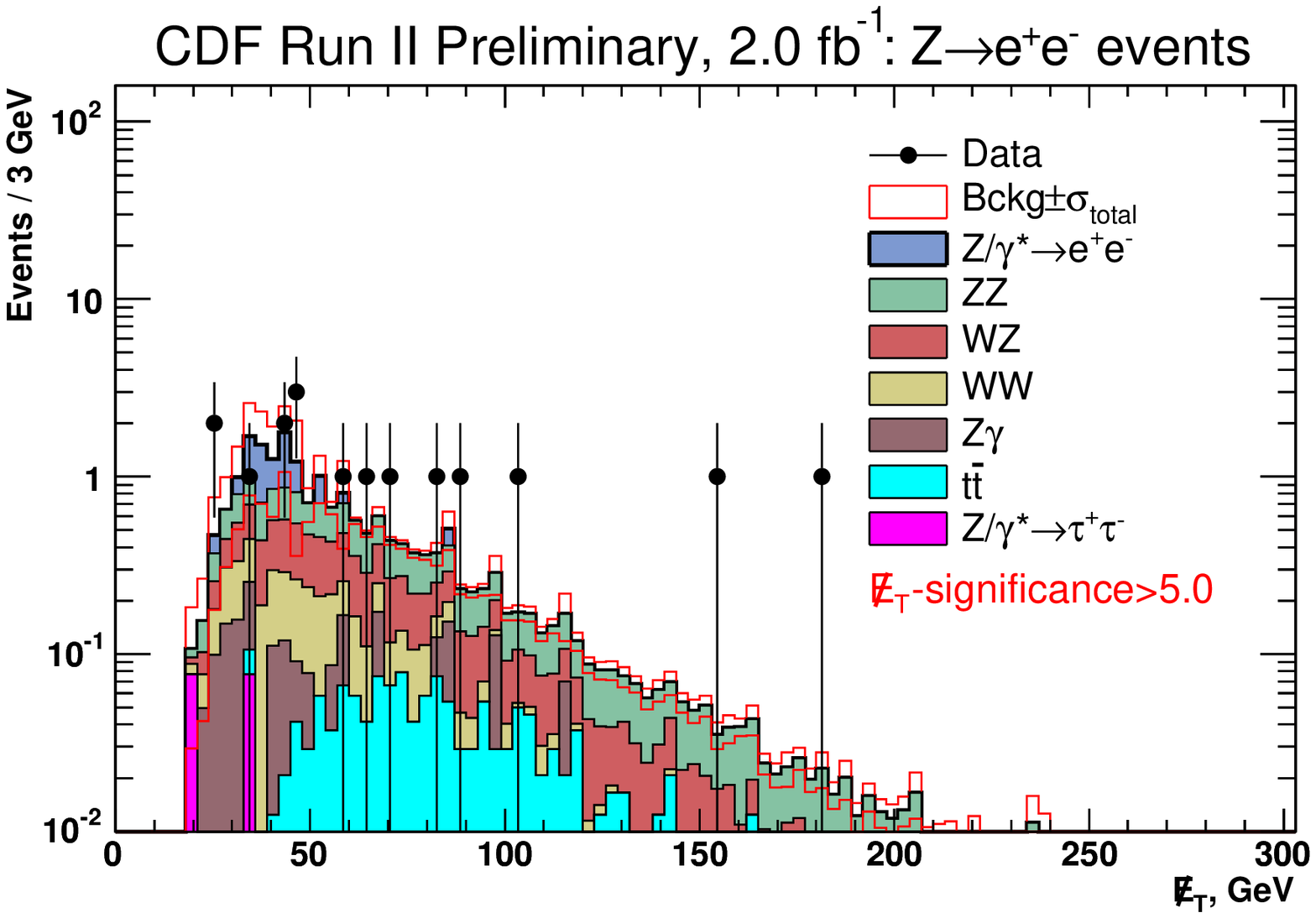}
\includegraphics[width=0.4\linewidth]{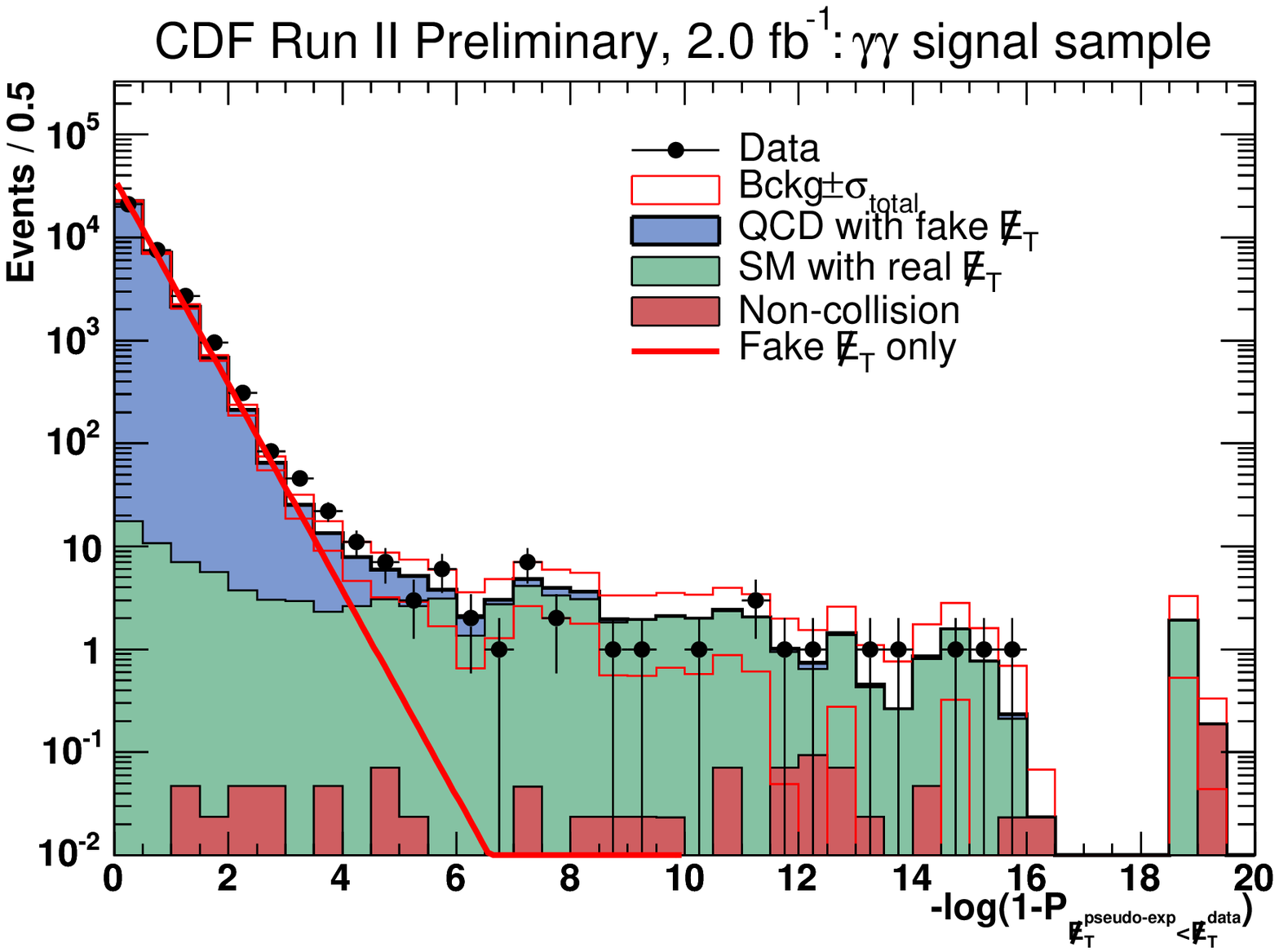}
\includegraphics[width=0.4\linewidth]{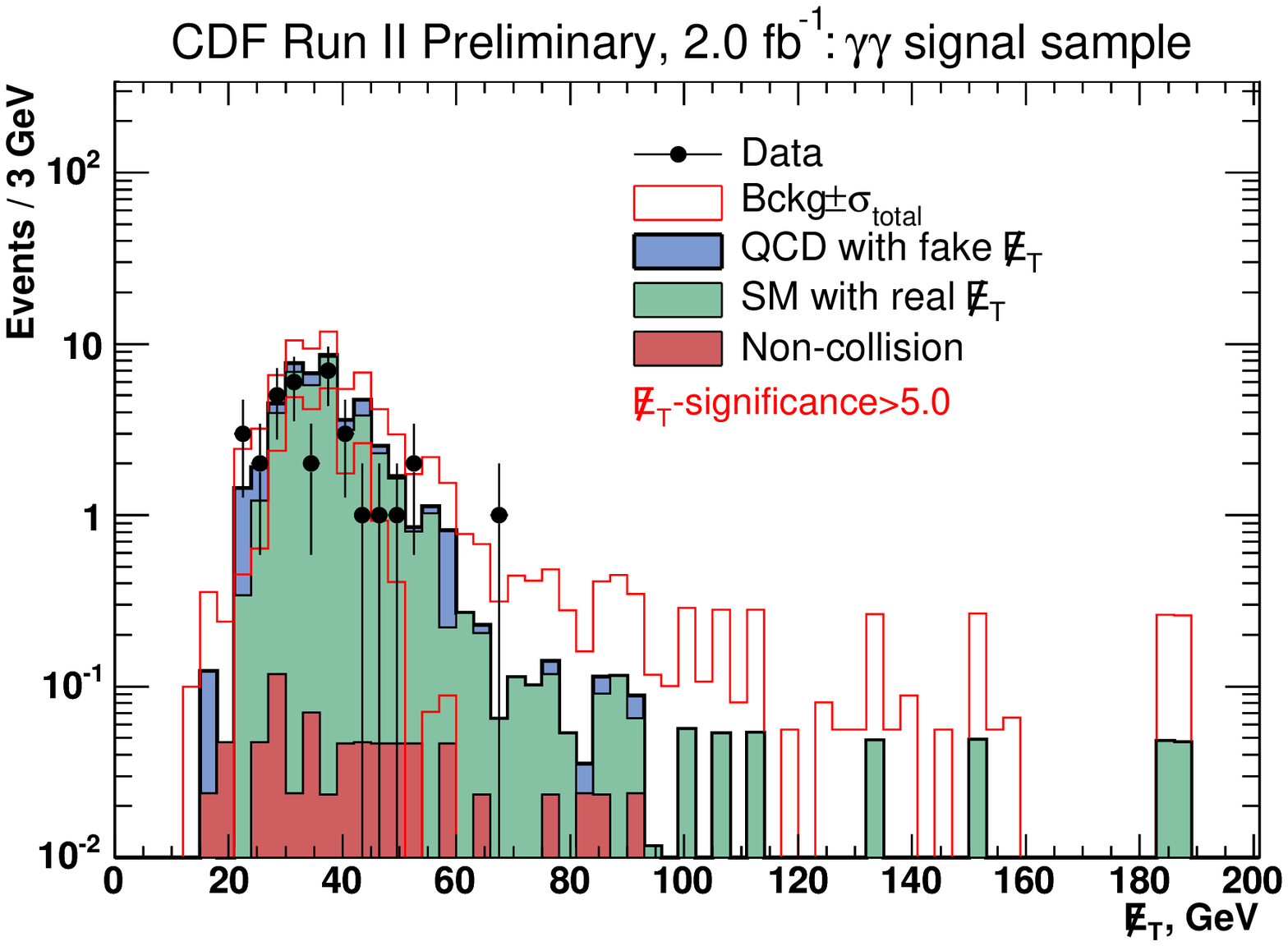}
\caption{Top: observed \mett\ distribution in $Z \rightarrow e^{+}e^{-}$ control sample
  and \mett\ distribution  after \mett\ significance~$ > 5$ cut.\\
  Bottom: \mett\ significance distribution in di-photon signal sample
  and observed \mett\ distribution after \mett\ significance~$ > 5$ cut.} 
\label{zeecs}
\end{figure*}

Two top plots in Figure~\ref{zeecs} show the model performance
for $Z\rightarrow e^{+}e^{-}$ events. After applying \mett\ significance cut 
only SM processes with real \mett\ are left in the sample.
The bottom two plots in Figure~\ref{zeecs} show the technique applied
to the sample of di-photon events. The observed number of events (34) is 
consistent with the sum of a predicted number of SM 
events ($48.6 \pm 7.5$). Two largest background contributions are electroweak
events with real \mett\ ($41.6 \pm 7$) and events with fake \mett\ ($6.2 \pm 2.7$).  

\section{Photon + jets + \mett}

\begin{figure*}[h]
\centering
\includegraphics[angle=90, width=0.4\linewidth]{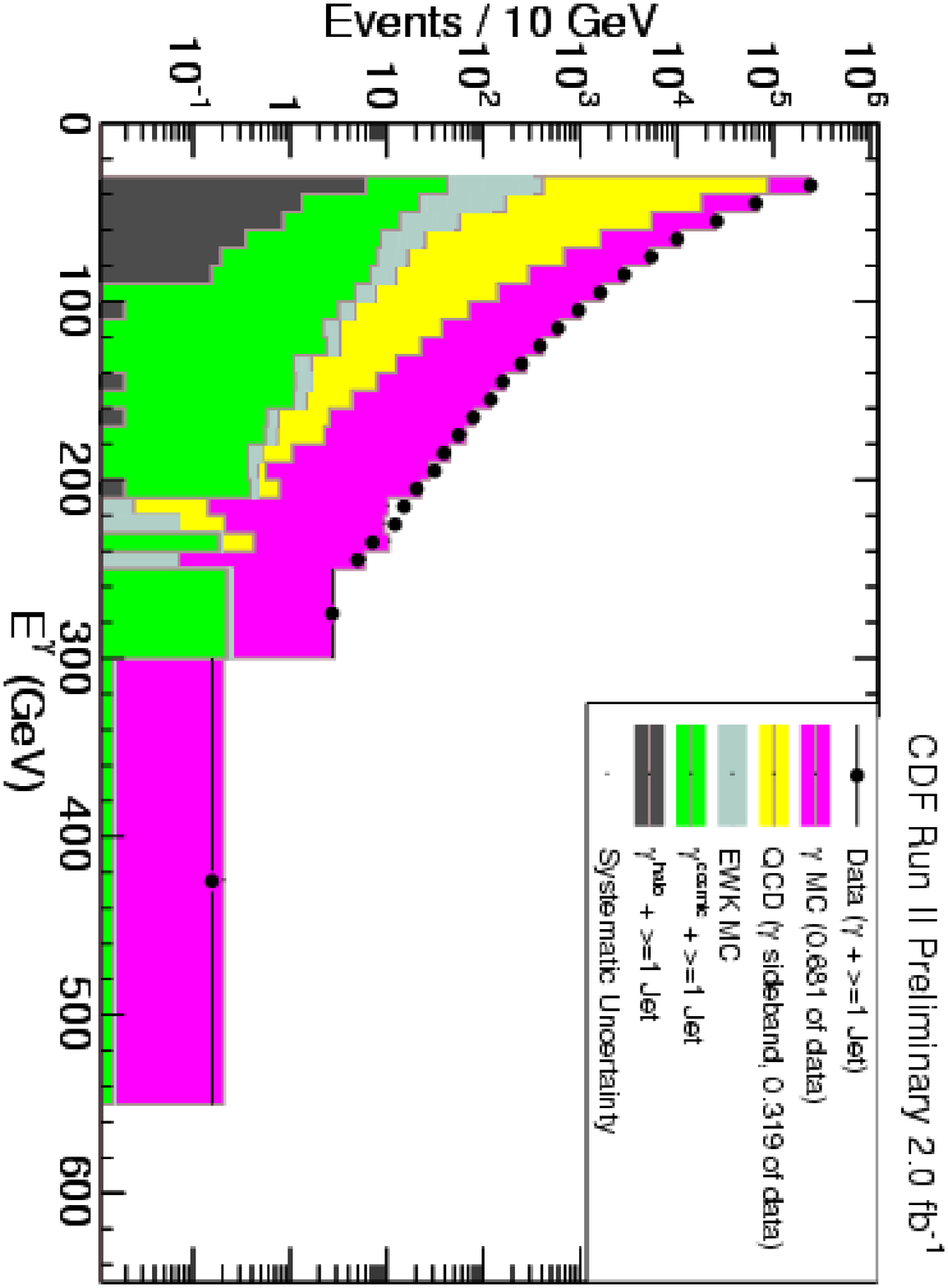}
\includegraphics[angle=90, width=0.4\linewidth]{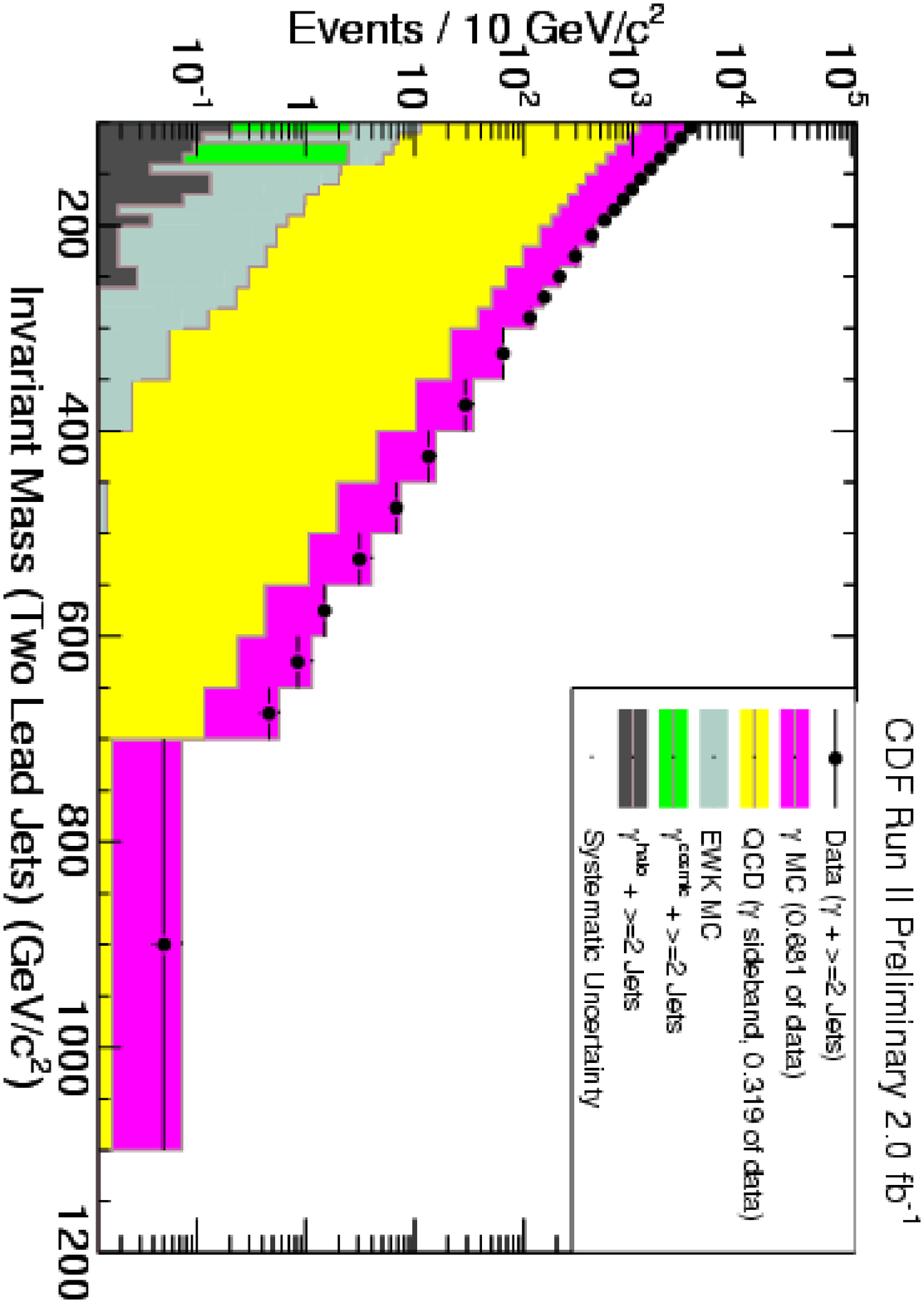}
\caption{Two kinematic distributions that compare data with the expected backgrounds
from SM. They show excellent agreement over several magnitudes of production cross-section} 
\label{g2jets1}
\end{figure*}

Many new physics models predict mechanisms that could produce a 
Photon~+~jets~+~\mett\ signature.  This section presents the generic search in 
the $\gamma$ + jets. A variety of techniques are applied to estimate 
the SM expectation and non-collision backgrounds. We examine
several kinematic distributions like the ones shown in Figure~\ref{g2jets1}.
At the moment the background estimation describes data well in the wide
kinematic regions that spans five orders of magnitude in production 
cross-section. The next step will be applying the \mett\ model described in
the previous section in order to select interesting events with large real 
\mett.

\section {Photon + b-jet + jet + \mett}

This section describes the results of a search for new physics in events containing
a photon, \mett, a b-tagged jet, and a second jet.  
This final state has low background from SM and is a promising place to look for new 
phenomena. Various extensions of the SM, such as GMSB, predict enhanced decay rates 
into this final state.

\begin{figure*}[h]
\centering
\includegraphics[width=0.4\linewidth]{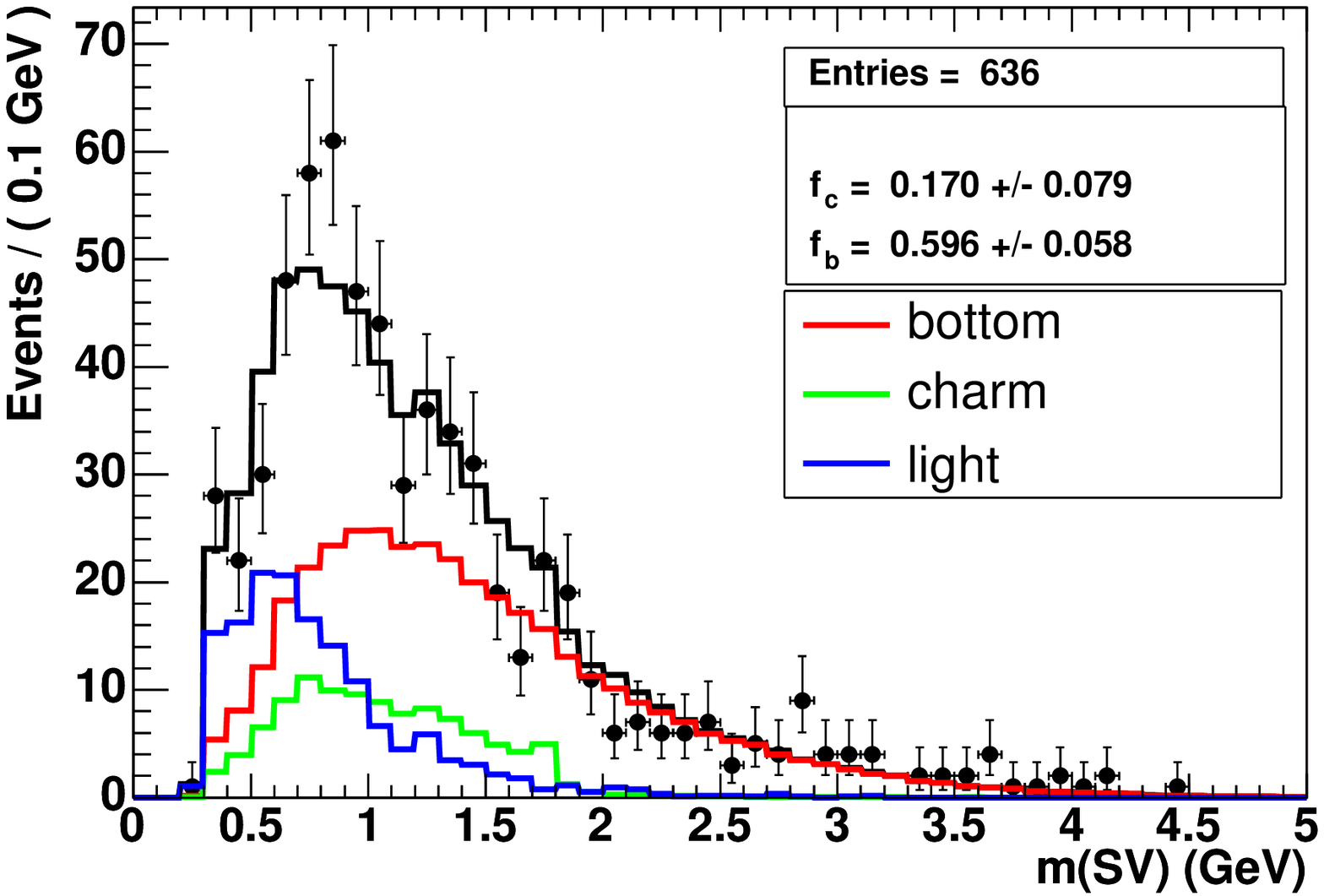}
\includegraphics[width=0.4\linewidth]{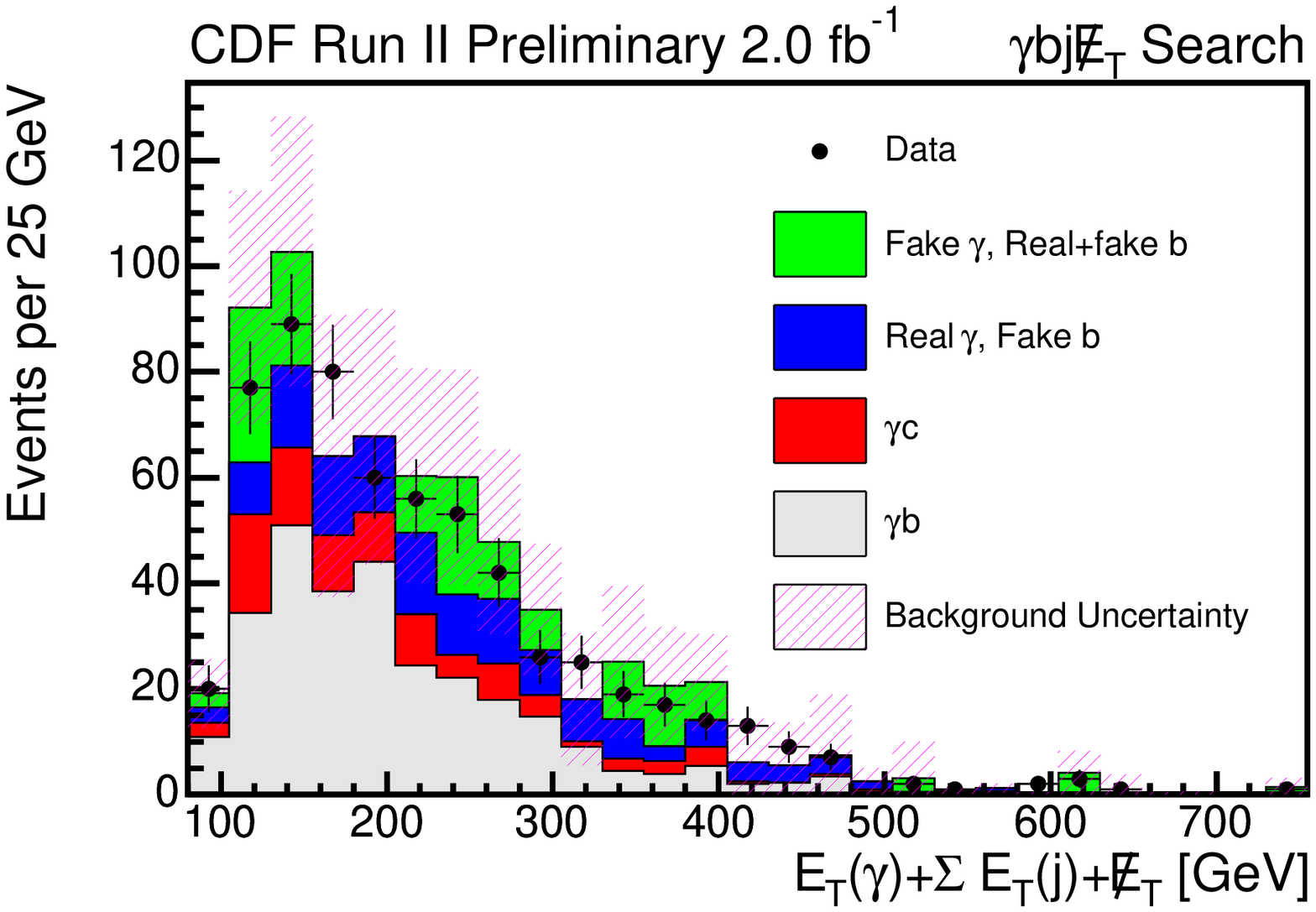}
\caption{The left plot shows the fit of Secondary vertex mass in the 
photon~+~b-jet~+~jet signal sample.\\
The right plot shows the scalar $P_{\rm T}$ sum of photon, all jets 
with $E_{\rm T} > 15$~GeV, $|\eta| < 2.0$, and \mett.} 
\label{gbj1}
\end{figure*}

We find $617$ events with $\mett > 25$~GeV and 
two jets with $E_{\rm T} > 15$~GeV, at least one identified as originating from 
a b quark, versus an expectation of $637 \pm 139$ events.
Figure~\ref{gbj1} shows the quark mass template fit used to estimate the backgrounds
from light quarks and the predicted $H_{\rm T}$ distribution. 

\section{Photon + b-jet + lepton + \mett}

This section presents a search for anomalous production of
$\gamma$~+~b-jet~+~lepton~+~\mett\ events. This signature occurs in GMSB models  
and other extensions of the SM.
Such models would produce this signature with large total energy $H_{\rm T}$.
The SM backgrounds are dominated by production of two gauge bosons, $W\gamma$, 
and two third-generation quarks ($t$ and $b$). We find 28 events versus a SM
expectation of $27.9 \pm 3.6$ events. 

A search for the production of top pairs with an additional photon, $t\bar{t} + \gamma$, 
is a natural extension of the signature-based search because 
$t\bar{t} + \gamma$ is characterized by a high $P_{\rm T}$ lepton, photon, 
b-tagged jet, and \mett. In addition we require large total transverse energy $H_{\rm T}$ 
and 3 or more jets so that radiative top-pair events dominate the SM predictions.
We observe 16 $t\bar{t} + \gamma$ candidate events versus an expectation of $11.1 \pm 2.3$ events. 
Assuming the difference between the observed number and the predicted non-top SM total is due 
to top production, we measure  $t\bar{t} + \gamma$ cross-section to be $0.15 \pm 0.08$~pb.
The probability, assuming no true   $t\bar{t} + \gamma$ SM 
signal, for the background alone to produce at least as many events (16) 
as observed in data, is 1\% ($2.3~\sigma$). Assuming SM   $t\bar{t}$ production, 
we estimate the  $t\bar{t} + \gamma$ cross-section to be $0.15 \pm 0.08$~pb. Theoretical 
cross-section $0.08 \pm 0.011$~pb is obtained from LO MadGraph cross-section 
(0.076~pb) multiplied by ${\rm k_{factor}} = 1.10 \pm 0.15$. We compare the 
measured $t\bar{t} + \gamma$ cross-section with other SM cross-sections.

\begin{figure*}[t]
\centering
\includegraphics[angle=90, width=0.8\linewidth]{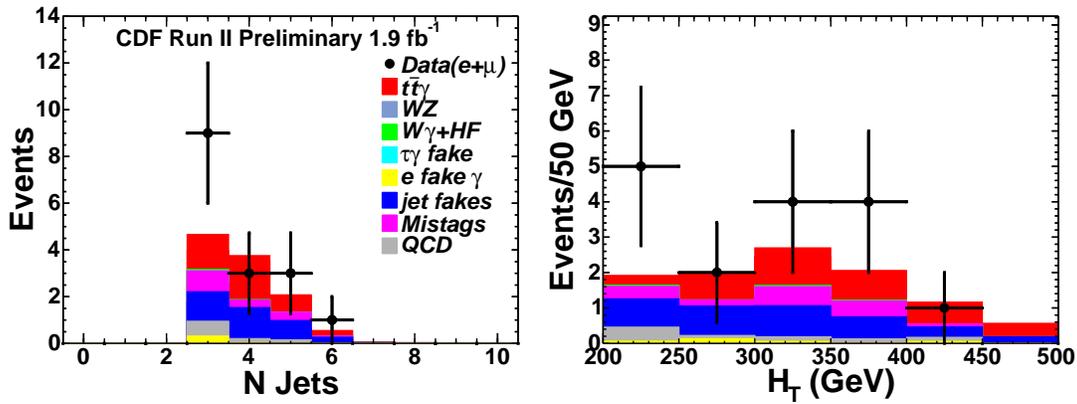}
\caption{The distributions in $H_{\rm T}$ and Number of Jets, observed in the 
$t\bar{t} + \gamma$ search. The histograms show the expected SM contributions and 
observed data.} 
\label{ttg2}
\end{figure*}

\begin{acknowledgments}
I would like to thank all CDF collaboration, especially people in the groups
that produced such interesting physics described in this article. I would
also like to thank the organizers for orchestrating such a wonderful conference.
 
\end{acknowledgments}


\begin{thebibliography}{9}
\bibitem{htdef} $H_{\rm T}$ is the scalar sum of the transverse energies 
of the reconstructed jets.
\end{thebibliography}
\end{document}